\documentclass[11pt]{article}

\usepackage[preprint]{acl}

\usepackage{times}
\usepackage{latexsym}

\usepackage[T1]{fontenc}
\usepackage[utf8]{inputenc}

\usepackage{microtype}

\usepackage{inconsolata}

\usepackage{graphicx}

\usepackage[most]{tcolorbox}
\usepackage{booktabs}
\usepackage{tabularx}
\title{SkillQR: Agentic Skill Discovery for Reason-Intensive Query Rewriting}
\newcommand{\emorymark}{\ensuremath{\diamondsuit}}
\newcommand{\sysumark}{\ensuremath{\clubsuit}}
\author{
  \textbf{Kenan Jiang}\textsuperscript{\emorymark} \quad
  \textbf{Meixiu Long}\textsuperscript{\sysumark} \quad
  \textbf{Zihao Zhang}\textsuperscript{\emorymark} \quad
  \textbf{Ran Xu}\textsuperscript{\emorymark} \quad
  \textbf{Eugene Agichtein}\textsuperscript{\emorymark} \quad
  \textbf{Fei Liu}\textsuperscript{\emorymark}
  \\[0.5em]
  \textsuperscript{\emorymark}Emory University
  \qquad
  \textsuperscript{\sysumark}Sun Yat-sen University
    \\[0.25em]
    \small
  \texttt{\{kenan.jiang, zihao.zhang, ran.xu, eugene.agichtein, fei.liu\}@emory.edu}
  \\
  \small
  \texttt{longmx7@mail2.sysu.edu.cn}
}

\begin{document}
\maketitle
\begin{abstract}
Dense retrievers provide grounded evidence that mitigates hallucinations in outputs from Large Language Models (LLMs). Real-world queries are often verbose, ambiguously framed, or require multi-step reasoning, introducing irrelevant details that weaken semantic matching for dense retrievers. Query rewriting is a widely adopted pre-retrieval technique that leverages LLMs to reformulate queries and bridge the semantic gap between such queries and the retrieval target. Existing rewriting approaches either rely on expensive proprietary models, substantial training data, or iterative retrieval and rewriting, while largely applying a uniform rewriting strategy regardless of the retrieval corpus. We observe that effective query rewriting instructions are not universal: the optimal strategy depends on the retrieval target, rather than the model alone. We introduce SkillQR, a query rewriting framework that automatically discovers reasoning strategy sets from the retrieval target and dynamically routes a query to the appropriate rewriting strategy at inference time. Without relying on training data, SkillQR achieves competitive performance compared to fine-tuned rewriting models on the BRIGHT and R2MED benchmarks, while also providing lower inference latency than iterative rewriting approaches. 
\end{abstract}

\section{Introduction}

Retrieval is a core component of modern search and retrieval-augmented generation (RAG) systems~\citep{karpukhin-etal-2020-dense,su-etal-2023-one,wang-etal-2024-improving-text, sengupta-etal-2026-tooldreamer}. Retrieval quality directly determines the evidence available for downstream generation~\citep{lewis2020retrieval,xu2024retrieval,xu2025acesearcher}. In real-world applications, however, user queries rarely appear in a retrieval-friendly form. Instead, they are often indirect, ambiguously framed, and expressed through surface context rather than the underlying information need~\citep{asai-etal-2023-task}. Such queries may contain irrelevant details or omit critical structure, which hinders effective retrieval. This challenge is especially pronounced in \emph{reason-intensive retrieval}, where successful retrieval requires inferring the latent reasoning that connects the query to relevant evidence~\citep{chen2025reasonembed,su2025bright,shao2025reasonir}. As a result, even strong dense retrievers can fail when the surface form of the query does not align with the retrieval target. 

Query rewriting~\citep{wang-etal-2023-query2doc, yao-etal-2025-expandr, ma-etal-2023-query} is a natural approach to this problem. The key idea is to transform the original query into a form that better reflects the information needed for retrieval, such as implicit concepts, intermediate reasoning steps, or structural constraints. Recent work shows that such reformulation can substantially improve retrieval for complex and reasoning-intensive queries, making rewriting a practical interface between users and retrievers~\citep{tongsearch,ReDI,lai-etal-2025-adacqr}. 
Despite these advantages, existing query rewriting systems remain limited in both \emph{efficiency} and \emph{adaptability}. Many approaches rely on expensive proprietary LLMs or on reinforcement learning (RL)-based training with substantial curated data~\citep{chen2025reasonembed,tongsearch}. Others depend on iterative retrieval-and-rewrite procedures or repeated corpus interaction, which further increases inference cost and latency~\citep{lei-etal-2025-thinkqe,li2025llm}. As a result, current methods can be costly to build, difficult to transfer across corpora, and impractical to deploy in privacy-sensitive or resource-constrained settings. 

Beyond these practical limitations, existing approaches largely keep the rewriting strategy fixed across retrieval settings, such as an instruction to "generate relevant knowledge for the query", with improvements coming mainly from stronger models or additional training. We argue that this is fundamentally incomplete: the optimal rewrite depends not only on the query, but also on the target corpus. For example, the query "how to implement binary search" requires fundamentally different rewrites when targeting theoretical computer science papers versus Python Stack Overflow answers. However, this dependence on the \emph{retrieval target} is largely overlooked in prior work. This motivates a corpus-aware view of query rewriting, in which \textbf{the rewriting strategy should adapt to the corpus rather than remain fixed}.

Together, these challenges highlight the need for a \textbf{lightweight and adaptive query rewriting framework} that improves retrieval through better strategy design rather than heavier models or iterative procedures. Motivated by this perspective, we propose SkillQR, a training-free query rewriting framework that automatically discovers corpus-grounded reasoning strategies and dynamically applies them at inference time. SkillQR extracts a compact set of rewriting strategies from the target corpus and routes each query to the appropriate strategy, enabling rewriting that is aligned with the retrieval target. 
Our contributions are summarized as follows:

\begin{itemize}
    \item We show that the rewriting strategy impacts retrieval performance, and that it should be conditioned on the 
    retrieval target.
    \item We propose \textbf{SkillQR}, a corpus-guided query rewriting framework that automatically discovers reusable rewriting skills without requiring relevance labels or model fine-tuning.
    \footnote{Our source code will be made publicly available upon acceptance.}
    \item We evaluate SkillQR against existing rewriting methods on two reason-intensive retrieval benchmarks, BRIGHT and R2MED, demonstrating performance comparable to previous methods using fewer resources.
\end{itemize}

\section{Related Work}
\subsection{Query Rewriting for Reason-Intensive IR}
The BRIGHT benchmark~\citep{su2025bright} evaluates reasoning-intensive retrieval across twelve subjects, ranging from programming to natural sciences. It shows that even state-of-the-art dense retrievers remain insufficient when the reasoning gap is not addressed. R2MED~\citep{li2025r2med} is a similar benchmark focusing on the medical domain content. Current research addresses reason-intensive retrieval in three ways: (1) Reasoning-aware retrievers, such as DIVER~\citep{long2025diver}, Reason-Embed~\citep{chen2025reasonembed}, and ReasonIR~\citep{shao2025reasonir}, which utilize LLMs to generate hard negatives for contrastive learning, training retrievers to inherently bridge the reasoning gap; and (2) Reranking, where specialized ranking models refine the relevance order of retrieved outputs~\citep{song-etal-2025-limrank, ji2024reasoningrank}. 
(3) Our work focuses on query rewriting as a complementary approach that can be combined with advances in both dense retrieval and reranking. Representative methods include TongSearch-QR~\citep{tongsearch} and Reasoner-Rewriter~\citep{chen2025reasonembed}, which use reinforcement learning or specialized fine-tuning to produce retrieval-effective rewrites.

Existing query rewriting systems suffer from several limitations. The basic approach relies on expensive proprietary models such as GPT-4o~\citep{su2025bright}, which hinders long-term deployment and raises privacy concerns when sensitive queries must be sent to third-party services. Training a customized query rewriter for complex query understanding is the predominant approach~\citep{tongsearch, chen2025reasonembed, ReDI}. While more cost-effective at inference time, these approaches demand substantial training resources and carefully curated datasets, often limiting their transferability across domains. Lastly, training-free methods such as ThinkQE~\citep{lei-etal-2025-thinkqe} employ pseudo-relevance feedback (PRF) by iteratively retrieving and adding top-ranked documents in LLM prompts to rewrite queries. This process increases online rewriting latency, as both the prompt length and generated output grow with each iteration, while also relying on strong reasoning LLMs, making it less practical in resource-constrained deployment settings. 
\begin{table*}[t]
\centering
\label{tab:oracle}
\setlength{\tabcolsep}{4.0pt}
\renewcommand{\arraystretch}{1.05}
\footnotesize
\begin{tabular}{lc
@{\hspace{6pt}}
ccccccc
@{\hspace{10pt}}
cc
@{\hspace{10pt}}
cccc}
\toprule
& &
\multicolumn{7}{c}{\textbf{StackExchange}} &
\multicolumn{2}{c}{\textbf{Coding}} &
\multicolumn{3}{c}{\textbf{Theorem-based}} \\
\cmidrule(r{6pt}){3-9} \cmidrule(r{6pt}){10-11} \cmidrule(r{6pt}l{6pt}){12-14}
\textbf{Model} & \textbf{Avg} &
\textbf{Bio.} & \textbf{Earth.} & \textbf{Econ.} & \textbf{Psy.} &
\textbf{Rob.} & \textbf{Stack.} & \textbf{Sus.} &
\textbf{Leet.} & \textbf{Pony} &
\textbf{Aops} & \textbf{TheoQ.} & \textbf{TheoT.} \\
\midrule
Original & 24.4 & 26.2 & 31.4 & 23.5 & 30.1 & 18.1 & 23.9& 21.0 & 35.0& 10.4& 14.9 & 31.6 & 26.7 \\
Generic & 30.4 & 38.4 & 42.4 & 28.8 & 38.9 & 25.8 & 28.5 & 27.4 & 31.7 & 15.3 & 9.7 & 40.2 & 37.5\\
Oracle & \textbf{33.0} & 38.4 & 42.4 & 28.8 & 38.9 & 25.8 & 28.5 & 27.4 & \textbf{37.2} & \textbf{30.1} & 10.8 & 43.1 & \textbf{45.0}\\
\bottomrule
\end{tabular}
\caption{NDCG@10 of different rewriting strategies on BRIGHT with ReasonIR Retriever~\cite{shao2025reasonir}. Oracle uses manually designed strategies on Coding and Theorem-based and improves retrieval on them. Generic uses a subject-agnostic rewrite strategy for all subjects. Major improvements are in Bold. All rewrites are generated with Qwen2.5-7B-Instruct.}
\label{tab:oracle}
\end{table*}

\subsection{Skills in LLM Agents}
Agent skills provide a modular way to manage and reuse instructions and procedures across tasks~\citep{anthropic2025skills,zhou2026comprehensive}. Voyager~\citep{wang2024voyager} demonstrates automatic skill discovery through environment interaction and feedback, with subsequent work exploring skill refinement from execution failures~\citep{alzubi2026evoskill, zheng2025skillweaver} and recurring user requirements~\citep{yang2026autoskill}. Together, these works highlight skills as a lightweight and flexible approach to agent adaptation that can reduce reliance on manually designed workflows. Motivated by these benefits, SkillQR combines offline corpus-guided skill discovery with query-dependent routing to adapt query rewriting to the target corpus without model fine-tuning.
\section{SkillQR}
\subsection{Motivation: Rewriting Strategy}
\label{sec:oracle}
We conduct an oracle experiment on the BRIGHT benchmark~\citep{su2025bright} to quantify the benefit of adopting corpus-grounded rewriting strategies. Results are shown in Table~\ref{tab:oracle}. Both \textbf{Generic} and \textbf{Oracle} use the same rewriting framework. However, \textbf{Generic} applies a generic strategy across all subjects, while \textbf{Oracle} employs human-curated, subject-specific strategies for Coding and Theorem-based tasks, and retains the generic strategy for StackExchange questions. We provide detailed human-designed strategies in Appendix~\ref{appendix:human}.\footnote{We adopt human and generic instructions from Reasoner-Rewriter~\citep{chen2025reasonembed}.}
\newtcolorbox{promptbox}[1]{
  colback=white,
  colframe=teal!70,
  coltitle=white,
  colbacktitle=teal!70,
  title=\textbf{#1},
  boxrule=0.2pt,
  left=6pt, right=6pt, top=4pt, bottom=4pt,
  fontupper=\small
}
\begin{promptbox}{Rewrite Instruction Template}
Given a task and an input, first analyze the task and the input, then present the complete long output...
{\scriptsize\textcolor{gray}{(Abbreviated for presentation)}}\\[2pt]
\textbf{Task:} \textit{\{rewrite strategy\}} \textbf{Input:} \textit{\{query\}}\\[4pt] 
\colorbox{gray!12}{\parbox{\dimexpr\linewidth-12pt\relax}{
\textbf{Generic Rewrite Strategy:} Generate information that is relevant and helpful to address the question in detail.}
}\\[2pt]
\colorbox{green!8}{\parbox{\dimexpr\linewidth-12pt\relax}{
\textbf{Human Strategy Example (Pony):} Generate the relevant \textcolor{blue}{Pony syntax and language features}, and provide a detailed explanation of their usage.
}}
\end{promptbox}

Table~\ref{tab:oracle} shows that human-curated strategies (oracle) improve average NDCG@10 from 30.4 to 33.0 over the generic baseline, a gain of +2.6 points achieved purely through strategy adaptation with no change to the underlying model. Notably, subject Leet, where generic rewriting previously hurt performance, shows a great performance recovery with the oracle instruction. On average, the oracle demonstrates that retrieval benefits from the rewriting strategy. However, manually crafting subject-specific strategies requires a holistic understanding of the retrieval corpus that is costly to obtain and does not scale. This motivates SkillQR's approach of recovering this signal automatically via LLM agent interactions.
\begin{figure*}[t]
    % \centering
    \includegraphics[width=1.0\textwidth]{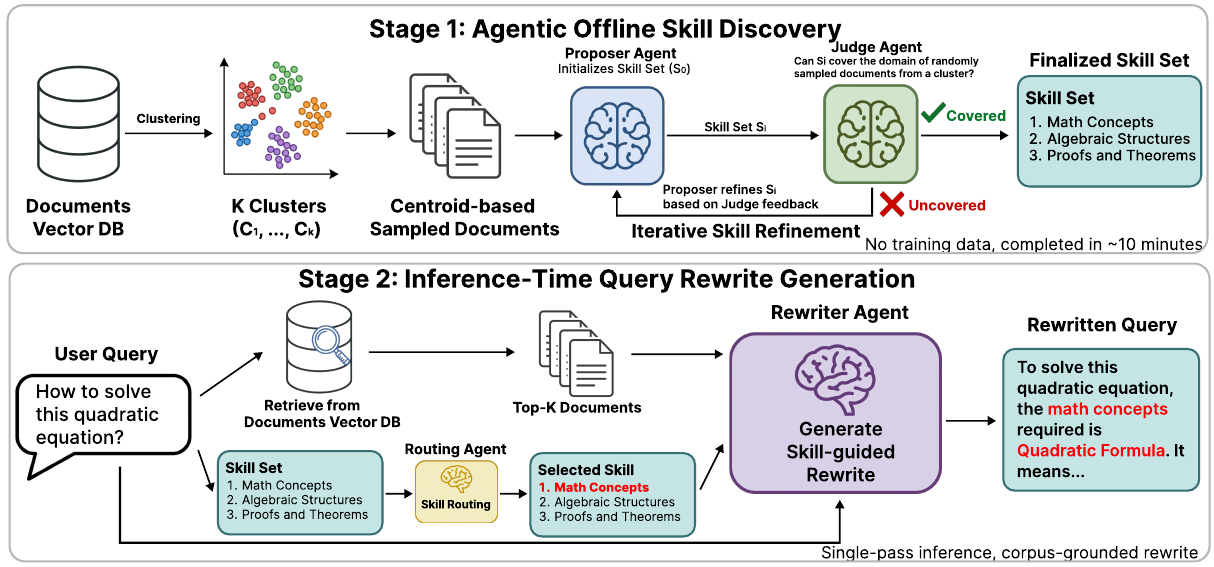}
    \vspace{-1.0em}
    \caption{SkillQR has two stages: (1) Agentic offline skill discovery mines rewriting skills from the document collection; (2) the router selects a query-grounded skill for the rewriter to rewrite at inference time.}
    \label{fig:SkillQR}
\end{figure*}
\subsection{Problem Formulation}
Given a query $q$ and a document collection $D$, let $D_q^{+} \subset D$ denote the small set of relevant documents to $q$. We posit that a reasoning-intensive query $q$ is connected to $D_q^{+}$ through a latent reasoning trajectory $T$. This trajectory captures the implicit reasoning required to bridge the gap between the query and the relevant documents (e.g., identifying core algorithms).

The retrieval system comprises a dense retriever $R$ and an LLM-based rewriter $W_{\text{LLM}}$. At inference time, the rewriter produces $q^* = W_{\text{LLM}}(q)$. The retriever $R$ retrieves with the updated $q^*$, based on cosine similarity, to obtain the documents $D_{q^*}^R \subset D$. The goal of $W_{\text{LLM}}$ is to find $q^*$ such that:
\[
\mathcal{M}(D_q^{+}, D_{q^*}^R) > \mathcal{M}(D_q^{+}, D_{q}^R)
\]
where $\mathcal{M}$ is a ranking metric such as NDCG@$10$.
\subsection{Agentic Offline Skill Discovery}
SkillQR discovers corpus-grounded rewriting strategies through an offline agentic pipeline operating over the document collection $D$, requiring no training data. We provide the detailed prompts of SkillQR in Appendix~\ref{appendix: skillqr_prompts}.

\paragraph{Corpus Exploration.} We use the document embeddings built by the dense retriever $R$, thereby avoiding any additional encoding cost. The document collection is partitioned into $K$ clusters $\{C_1, \ldots, C_K\}$, where $K \approx \sqrt{|D|}$, using a clustering algorithm in the embedding space. For each cluster $C_k$, we select a small set of representative documents $\tilde{C}_k \subset C_k$ closest to the cluster centroid. A proposer agent $P_{\text{LLM}}$ characterizes each cluster using two attributes: a \textit{domain label} and a \textit{content-type} descriptor (e.g., domain: math, content type: proofs). This results in a set of $K$ cluster characterizations, denoted as $\mathcal{A} = \{(d_1, t_1), \ldots, (d_K, t_K)\}$, where each pair corresponds to the domain and content type of a cluster.

\paragraph{Skill Initialization.}
We define \textbf{skills as a set of reasoning strategies discovered from the retrieval corpus}, which are stored in the rewriter’s memory and selectively applied at inference time. An LLM-based proposer $P_{\text{LLM}}$ constructs an initial set of rewriting instructions $S_0$ from the cluster characterizations $\mathcal{A}$. It begins with the generic rewrite strategy in Section~\ref{sec:oracle}, denoted $\mathcal{I}$, which requires no domain expertise to craft.
\begin{promptbox}{Skill Initialization Prompt}
Here are the groups in a collection: \{$\mathcal{A}$\}. A naive writer uses this generic skill: \{$\mathcal{I}$\}. This is too generic. Expand it to the \textbf{minimal} set of skills that covers all groups. {\scriptsize\textcolor{gray}{(Abbreviated for presentation)}}
\end{promptbox}
\paragraph{Iterative Skill Refinement.}
A judge LLM $J_{\text{LLM}}$ validates the proposed skill set $\mathcal{S}_0$ through an iterative refinement loop. In each round $i$ starting 0, the judge evaluates $\mathcal{S}_i$ against each cluster $C_k$ by randomly sampling a subset of documents $\hat{C}_k \subset C_k$.

For each $\hat{C}_k$, the judge assesses whether any $s \in \mathcal{S}_i$ can guide an LLM to generate texts that are similar to documents in $\hat{C}_k$. If $J_{\text{LLM}}$ finds such skill $s' \in \mathcal{S}_i$, then $\mathcal{S}_i$ covers $\hat{C}_k$. Otherwise, it specifies the \textit{missing content-type} $\delta_k$ required for coverage. Formally, for each cluster $C_k$, the judge outputs a coverage decision:
\[
J_{\text{LLM}}(\hat{C}_k, \mathcal{S}_i) \rightarrow 
\begin{cases}
(\textsc{covered}, s'_k) & s'_k \in \mathcal{S}_i \\
(\textsc{uncovered}, \delta_k) \\
\end{cases}
\]

For each uncovered cluster, the judge recommends in $\delta_k$ whether it can be addressed by refining an existing skill or requires the creation of a new one. Clusters assigned to existing skills trigger targeted refinement, where the corresponding skill is broadened to cover the missing knowledge while preserving its original domain focus. Clusters deemed novel are grouped together, and the proposer generates a new skill to cover them.
\begin{promptbox}{Judge-Proposer Interaction Example}
\textbf{Judge:} $\hat{C}_1$\,\textcolor{olive}{\texttt{COVERED}} by Skill 2 \;\textbar\; $\hat{C}_2$\,\textcolor{red}{\texttt{UNCOVERED}} $\delta_2$: \textit{climate dynamics} $\approx$ Skill 1\;\textbar\; $\hat{C}_3$\,\textcolor{red}{\texttt{UNCOVERED}} $\delta_3$: \textit{UI/UX design} $\not\approx$ any skill\\[3pt]
\textbf{Proposer:} $\hat{C}_2$ \textit{climate dynamics} $\rightarrow$ \textsc{refine} $s_1$ \;\textbar\; $\hat{C}_3$ \textit{UX design} $\rightarrow$ \textsc{new} $s_3$
{\scriptsize\textcolor{gray}{\hfill(Abbreviated for presentation)}}
\end{promptbox}

Uncovered clusters are revisited in subsequent rounds and reassigned to updated skills when possible. The refinement loop terminates when all clusters are covered or a maximum number of iterations $M$ is reached. The final skill set at termination, together with the generic instruction $\mathcal{I}$, is denoted as $\mathcal{S}_f$, which approximates the space of possible reasoning trajectories $T$ linking queries to their retrieval targets. $\mathcal{S}_f$ serves as the final skill set for the retrieval corpus $D$.
\subsection{Inference-Time Rewrite Generation}
\label{sec:skillqr_inference}
At inference time, rewrite agent $W_{LLM}$ first routes the query by selecting the most appropriate skill $s^* \in \mathcal{S}_f$ for query $q$ by evaluating the match between
the query's domain and each skill in $S_f$. The selected skill is then used to prompt the rewriter $W_{\text{LLM}}$ for query expansion. We provide the full routing and generation prompt in Appendix~\ref{appendix: skillqr_prompts}.

We first perform an initial retrieval using the original query $q$ with the dense retriever $R$, obtaining a set of top-ranked documents $D_R(q)$. These documents are used as pseudo-relevance feedback (PRF) for the rewriter. Conditioned on the selected skill $s^*$ and the PRF documents $D_R(q)$, the rewriter $W_{\text{LLM}}$ generates an expanded query guided by skill $s^*$. This process aims to articulate the latent reasoning trajectory $T$ connecting the query to relevant documents. The final query is constructed as:
\[
q^* = concat(q, W_{\text{LLM}}(q | D_R(q), \pi(s^*)))
\]
where $\pi(s^*)$ denotes the prompt instantiated with the selected
skill.
\section{Experiments}
\subsection{Datasets and Baselines}
We evaluate SkillQR on the BRIGHT~\citep{su2025bright} and R2MED~\citep{li2025r2med} benchmarks, reporting averaged and per-subject NDCG@10. 

We compare against the following baselines: (1) ThinkQE~\citep{lei-etal-2025-thinkqe} is an answer-driven query rewriting method built on QWEN-R1-Distill-14B~\citep{deepseekai2025deepseekr1incentivizingreasoningcapability}, a foundation model with enhanced thinking ability. It iteratively alternates between retrieval and rewriting for three rounds. At each rewriting step, ThinkQE uses the top-5 retrieved documents as PRF to generate the answer for the query. Following the upgrade in Diver-QExpand~\citep{long2025diver}, we replace ThinkQE's original BM25 retriever with the dense retriever that the rewriting system supports to obtain the PRF documents. (2) TongSearch-QR~\citep{tongsearch}, a Qwen2.5-7B-Instruct-based~\cite{Yang2024Qwen25TR} query rewriter finetuned with RL on 30k curated data samples. (3) Reasoner-Rewriter (RRewriter)~\citep{chen2025reasonembed}, a Qwen-based rewriter also finetuned with RL, but on 68k samples curated directly from BRIGHT documents.

To assess generalization, we evaluate all query rewriting methods on two retrievers: Diver-Retriever~\citep{long2025diver} and Reason-Embed~\citep {chen2025reasonembed}, both of which are leading dense retrievers on the reason-intensive retrieval tasks. Model and prompt details appear in Appendix~\ref{appendix:main_experiment}. 

\subsection{SkillQR Implementation Details}
We employ QWEN-R1-Distill-14B~\citep{deepseekai2025deepseekr1incentivizingreasoningcapability} to create proposer and judge agents for all offline skill discovery processes. Skills are discovered from embeddings generated by the system’s retriever (e.g., Reason-Embed or Diver-Retriever). In SkillQR, we apply MiniBatchKMeans~\citep{JMLR:v12:pedregosa11a} to partition each subject-specific retrieval corpus into $K$ clusters, where $K$ is dynamically set to approximately the square root of the corpus size. We then discover skills using at most $M=3$ refinement rounds and at most $N_S=5$ skills. We set 3 rounds to bound offline compute cost, and cap 5 skills to ensure the skill set remains compact so routing errors reflect skill quality rather than context overload in the router. We evaluate sensitivity to the cluster count $K$ and skill cap $N_S$ in Appendix~\ref{app:cluster_ablation}. Skill discovery time is a one-time offline cost. The full BRIGHT benchmark is mined in approximately 2 hours on a single H100 GPU, with the largest subject LeetCode (400k documents) completing in 21 minutes (Appendix~\ref{app:time_analysis}). At inference time, SkillQR uses a single LLM, either Qwen2.5-7B-Instruct or QWEN-R1-Distill-14B, for both skill routing and query rewriting, requiring no additional model components beyond the chosen backbone. As in ThinkQE, we use the top-5 documents retrieved with the original query as PRF for query rewriting.
\subsection{BRIGHT Results}
\begin{table*}[h]
\centering
\label{tab:sheet_style}
\setlength{\tabcolsep}{4.0 pt}
\renewcommand{\arraystretch}{1.05}
\footnotesize
\begin{tabular}{lccccccccccccccc}
\toprule
& & &
\multicolumn{7}{c}{\textbf{StackExchange}} &
\multicolumn{2}{c}{\textbf{Coding}} &
\multicolumn{3}{c}{\textbf{Theorem-based}} \\
\cmidrule(r{6pt}){4-10} \cmidrule(r{6pt}){11-12} \cmidrule(r{6pt}l{6pt}){13-15}
\textbf{Method} & \textbf{Size} & \textbf{Avg} &
\textbf{Bio.} & \textbf{Earth.} & \textbf{Econ.} & \textbf{Psy.} &
\textbf{Rob.} & \textbf{Stack.} & \textbf{Sus.} &
\textbf{Leet.} & \textbf{Pony} & \textbf{Aops} &
\textbf{TheoQ.} & \textbf{TheoT.}\\
\midrule
\multicolumn{15}{c}{\textit{Diver-Retriever}} \\
\midrule
Original & - & 31.9 & 48.9 & 50.0 & 32.0 & 41.9 & 30.2 & 28.0 & 33.6 & 13.4 & 16.1 & 8.7 & 37.5 & 42.6 \\
ThinkQE & 14B & 34.1 & 53.1 & 55.7 & 34.9 & 46.3 & 33.3 & 32.9 & 35.9 & 12.5 & 10.5 & 10.6 & 38.9 & 44.6\\
SkillQR & 14B & 34.0$^\dag$ & 54.3 & 54.9 & 34.7 & 45.7 & 33.1 & 32.0 & 36.7 & 10.3 & 11.6 & 9.1 & 40.4 & 44.8 \\
TongSearch & 7B & 32.6 & 50.6 & 51.8 & 31.9 & 46.2 & 29.7 & 28.9 & 33.6 & 13.8 & 12.5 & 10.9 & 39.2 & 42.1\\
RRewriter & 7B& 34.5 & 47.5 & 54.8 & 33.7 & 43.7 & 30.0 & 28.0 & 37.2 & 11.3 &30.2 & 8.9 & 41.1 & 47.8\\
SkillQR & 7B& 34.7$^\dag$ & 51.9 & 52.6 & 32.5 & 47.3 & 33.5 & 31.0 & 37.4 & 11.8 & 22.8 & 9.5 & 39.8 & 46.9 \\
\midrule
\multicolumn{15}{c}{\textit{Reason-Embed}} \\
\midrule
Original & - & 36.2 & 51.8 & 52.5 & 33.5 & 44.8 & 34.3 & 34.6 & 35.7 & 38.2 & 9.3 & 12.8 & 40.8 & 46.3 \\
ThinkQE & 14B & 39.3 & 61.4 & 55.8 & 33.6 & 49.2 & 32.9 & 40.7 & 40.3 & 34.2 & 16.5 & 15.1 & 42.9 & 48.6 \\
SkillQR & 14B & 38.5$^\dag$ & 55.9 & 55.0 & 35.1 & 48.5 & 35.5 & 38.1 & 38.0 & 35.9 & 15.6 & 14.6 & 43.2 & 47.2\\
TongSearch & 7B & 38.1 & 59.1 & 59.2 & 31.1 & 47.5 & 31.0 & 36.9 & 36.8 & 31.6 & 20.5 & 15.2 & 42.1 & 46.1\\
RRewriter & 7B & 38.4 & 55.1 & 59.2 & 32.9 & 44.5 & 32.0 & 37.6 & 35.0 & 31.4 & 23.5 & 16.2 & 45.2 & 48.2\\
SkillQR & 7B & 38.3$^\dag$ & 57.7 & 55.7 & 33.0 & 47.9 & 35.3 & 38.5 & 36.5 & 34.8 & 19.2 & 12.8 & 42.2 & 46.0\\
\bottomrule
\end{tabular}
\caption{NDCG@10 comparison of rewriting methods on BRIGHT. $^\dag$ indicates \textbf{no statistically significant difference} from the baselines of the same size (paired two-tailed t-test, p $>$ 0.05).}
\label{tab:ndcg_bright}
\end{table*}
\begin{table}[h]
\centering
\small
\setlength{\tabcolsep}{4pt}
\begin{tabular}{lccc}
\toprule
\textbf{Method} & \textbf{Train Data} & \textbf{Size} & \textbf{Time (s)} \\
\midrule
TongSearch  & 30k & 7B & $17.7 \pm 15.2$ \\
RRewriter   & 68k &7B & $21.1 \pm 8.8$ \\
SkillQR   & -- & 7B & $13.9 \pm 4.6$ \\
ThinkQE    & -- & 14B & $85.3 \pm 107.1$\\
SkillQR  & -- & 14B & $38.1 \pm 22.9$ \\
\bottomrule
\end{tabular}
\vspace{-0.5em}
\caption{Per-query rewrite time on BRIGHT with Diver-Retriever, averaged over 240 queries (20 per subject) on a single H100 GPU.}
\label{tab:inference_time}
\end{table}

Table~\ref{tab:ndcg_bright} reports SkillQR's performance on BRIGHT. At 14B-scale, SkillQR matches ThinkQE on all retrievers, reaching 34.0 and 38.5 NDCG@10 against ThinkQE's 34.1 and 39.3 on Diver and Reason-Embed, respectively. Despite comparable performance, SkillQR is approximately 45 seconds faster per query than ThinkQE, as shown in Table~\ref{tab:inference_time}. This efficiency gain comes from SkillQR's single-pass skill-guided rewriting compared to ThinkQE's three iterative answer-generation passes.

At 7B-scale, both TongSearch and RRewriter are explicitly finetuned on training data curated with BRIGHT as the target benchmark, and RRewriter further incorporates human expert skills, introduced in Section~\ref{sec:oracle}, in its rewrite instructions. Despite this, SkillQR achieves comparable performance, reaching 38.3 on Reason-Embed against RRewriter's 38.4, while requiring less inference time than both finetuned counterparts. We attribute the inference advantage to output token length: finetuned rewriters tend to follow reasoning patterns observed in training data, producing redundant tokens with diminishing retrieval value when a shorter rewrite could suffice. In summary, \textbf{SkillQR's corpus-grounded skill discovery matches methods that invest heavily in domain-specific data curation and RL-based finetuning, while requiring no training data, no reward design, and lower inference cost}.

The variance of the inference time from Table~\ref{tab:inference_time} is also informative. We observe a large variance in ThinkQE's inference time, arising from its answer-driven strategy combined with a strong thinking model. Subjects with inherently longer answers, such as AoPS (math) and LeetCode, trigger substantially more chain-of-thought generation than subjects like psychology. In contrast, SkillQR exhibits significantly lower variance at the 14B-scale, as the corpus-grounded skill provides a focused rewriting target, bounding the generation regardless of subject complexity. We provide additional time analyses in Appendix~\ref{app:time_analysis}. 

\subsection{R2MED for Performance Transferability}
\begin{table*}[t]
\centering
\setlength{\tabcolsep}{4.0 pt}
\renewcommand{\arraystretch}{1.05}
\footnotesize
\begin{tabular}{lcccccccccc}
\toprule
\textbf{Method} & \textbf{Size} & \textbf{Avg} &
\textbf{Bioinfo.} & \textbf{Bio.} & \textbf{IIYi.} & \textbf{Medical.} &
\textbf{MedQA.} & \textbf{MedXpert.} & \textbf{PMC-C.} &
\textbf{PMC-T.} \\
\midrule
\multicolumn{11}{c}{\textit{Diver-Retriever}} \\
\midrule
Original & - & 42.7 & 60.3 & 48.9 & 31.3 & 57.6 & 22.5 & 18.5 & 44.0 & 58.5 \\
ThinkQE & 14B & 49.3 & 69.2 & 54.7 & 33.8 & 61.7 & 36.3 & 23.0 & 51.0 & 64.6 \\
SkillQR & 14B & 48.9$^\dag$ & 70.1 & 54.1 & 36.0 & 62.7 & 32.9 & 26.5 & 44.9 & 64.3 \\
TongSearch & 7B & 44.5 & 62.8 & 50.6 & 33.3 & 61.8 & 26.2 & 21.1 & 41.0 & 59.0\\
RRewriter & 7B & 44.9 & 64.8 & 47.5 & 31.7 & 62.0 & 28.1 & 23.9 & 40.8 & 60.2 \\
SkillQR & 7B & 46.9$^*$ & 63.5 & 50.3 & 35.5 & 63.4 & 29.5 & 23.0 & 46.1 & 64.0\\
\midrule
\multicolumn{11}{c}{\textit{Reason-Embed}} \\
\midrule
Original & - &  42.0 & 62.8 & 51.8 & 22.1 & 61.7 & 25.6 & 21.8 & 38.1 & 51.6\\
ThinkQE & 14B & 49.1 & 67.7 & 59.8 & 30.9 & 67.8 & 37.7 & 30.4 & 45.6 & 53.2\\
SkillQR & 14B & 49.3$^\dag$ & 70.2 & 56.6 & 34.4 & 65.5 & 34.5 & 31.8 & 46.1 & 55.4 \\
TongSearch &7B& 43.7 & 62.2 & 59.1 & 24.3 & 64.4 & 28.5 & 25.3 & 36.7 & 48.7 \\
RRewriter &7B& 42.3 & 63.5 & 55.1 & 21.7 & 64.0 & 29.2 & 25.7 & 32.7 & 46.7 \\
SkillQR &7B& 46.1$^*$ & 68.8 & 58.3 & 26.1 & 64.8 & 30.6 & 27.1 & 37.8 & 55.4 \\
\bottomrule
\end{tabular}
\caption{NDCG@10 comparison of rewriting methods on the R2MED benchmark. $^\dag$ indicates \textbf{no statistically significant difference} from ThinkQE (paired two-tailed t-test, p > 0.05). $^*$ indicates \textbf{statistically significant difference} from TongSearch and RRewriter (paired two-tailed t-test, p < 0.05).}
\label{tab:ndcg_r2med}
\end{table*}
\begin{table*}[t]
\centering
\label{tab:sheet_style}
\setlength{\tabcolsep}{4.0 pt}
\renewcommand{\arraystretch}{1.05}
\footnotesize
\begin{tabular}{lccccccccccccc}
\toprule
& &
\multicolumn{7}{c}{\textbf{StackExchange}} &
\multicolumn{2}{c}{\textbf{Coding}} &
\multicolumn{3}{c}{\textbf{Theorem-based}} \\
\cmidrule(r{6pt}){3-9} \cmidrule(r{6pt}){10-11} \cmidrule(r{6pt}l{6pt}){12-14}
\textbf{Method} & \textbf{Avg} &
\textbf{Bio.} & \textbf{Earth.} & \textbf{Econ.} & \textbf{Psy.} &
\textbf{Rob.} & \textbf{Stack.} & \textbf{Sus.} &
\textbf{Leet.} & \textbf{Pony} & \textbf{Aops} &
\textbf{TheoQ.} & \textbf{TheoT.} \\
\midrule
Original & 31.9 & 48.9 & 50.0 & 32.0 & 41.9 & 30.2 & 28.0 & 33.6 & 13.4 & 16.1 & 8.7 & 37.5 & 42.6 \\
Generic &32.1& 48.0 & 51.5 &30.7 &43.1 & 28.3 & 28.8 & 36.4 & 12.4 & 18.4 &9.3&37.9&40.4\\
Human &33.7& 48.0 & 51.5 &30.7 &43.1 & 28.3 & 28.8 & 36.4& 12.3&29.8&10.0&	39.8 & 45.2 \\
SkillQR & 33.7 & 48.5 & 53.6 & 31.7 & 43.6 & 30.2 & 29.6 & 33.4 & 13.4 & 27.0 & 9.6 & 38.3 & 45.2 \\
\bottomrule
\end{tabular}
\caption{Skill quality NDCG@10 comparison on BRIGHT with Diver-Retriever and Qwen2.5-7B-Instruct rewriter. SkillQR in this table does not inherently include PRF. SkillQR matches human expert skills on average.}
\label{tab:skill_ablation}
\end{table*}
Table~\ref{tab:ndcg_r2med} reports SkillQR's performance on R2MED, evaluating its transferability to the medical domain. When using the 14B thinking-enhanced model backbone, SkillQR achieves statistically comparable performance on both retrievers with less inference cost, reaching 48.9 and 49.3 NDCG@10 on Diver and Reason-Embed, respectively, against ThinkQE's 49.3 and 49.1. This confirms that SkillQR's efficiency advantage on BRIGHT transfers to the medical domain without sacrificing retrieval quality, indicating that applying corpus-grounded skills can substitute for iterative refinement even when the corpus distribution shifts substantially.

At 7B-scale, SkillQR achieves \textbf{statistically significant improvements} over TongSearch and RRewriter on two retrievers, achieving a 46.9 NDCG@10 with Diver against RRewriter's 44.9. These baselines are fine-tuned on data curated with BRIGHT as their target benchmark, so the rewriters mainly learn to fill reasoning gaps present in BRIGHT's domains. When the target corpus shifts to R2MED’s medical domain, the learned rewriting patterns transfer poorly to the new retrieval target. SkillQR avoids this failure mode by design. Rather than baking a fixed rewriting strategy into the model weights, it mines skills from whichever corpus it is deployed against, so the rewriting strategies are tailored to the current retrieval target. \textbf{SkillQR’s adaptability in the medical domain supports our central claim that rewriting strategies should adapt to the retrieval corpus rather than remain fixed, while demonstrating that such adaptation can be achieved through a lightweight, training-free framework.}
SkillQR also maintains consistent retrieval improvement over the original query across Diver-Retriever and Reason-Embed, suggesting that corpus-guidance is a useful signal across dense retrievers.
\section{Discussions}
\subsection{Skill Effectiveness}
\label{sec:skill_effectiveness_ablation}
Table~\ref{tab:skill_ablation} investigates the quality of automatically discovered skills. We remove the PRF documents from the rewriter's input to better isolate the effect of the skill applied. On average, SkillQR outperforms the generic rewriting strategy, yielding a 1.6 gain and matching the human-designed skills. This confirms that the discovered skills are effective and may reduce the need for manual, holistic corpus inspection. Notably, the human-designed skills default to the generic instruction for all StackExchange subjects, implicitly assuming that domain-specific strategies are unnecessary for these topics. SkillQR discovers distinct strategies across all subjects, and the resulting improvements on subsets such as Rob (+1.9) suggest that the human's assumption is incorrect. This demonstrates that SkillQR can uncover strategies overlooked by human experts. The skill effectiveness also transfers to the Reason-Embed retriever in Appendix~\ref{app: skill_effectiveness}.
\begin{figure*}[h]
    \includegraphics[width=1.0\textwidth]{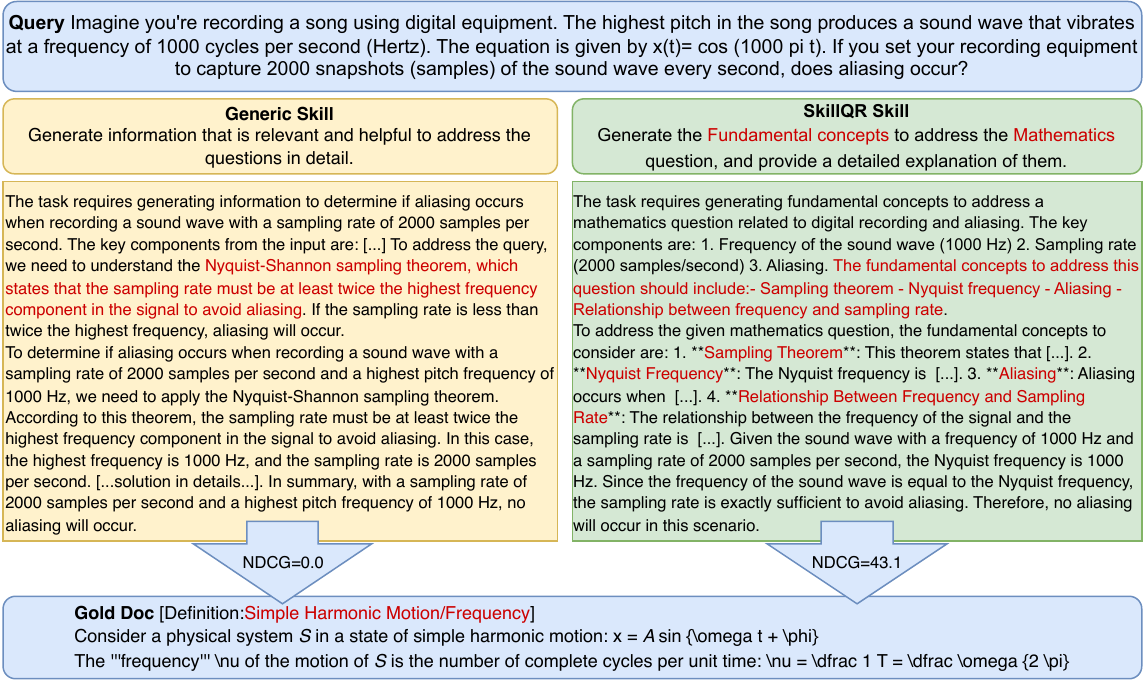}
    \vspace{-1.0em}
    \caption{SkillQR directs the focus of the rewrite from problem-solving into theorem-explaining, improving the NDCG@10 score from 0.0 to 43.1. Complete rewrites are in Appendix~\ref{sec:additional_qualitative}.}
    \label{fig:skillqr_case}
\end{figure*}
\subsection{Router Analysis}
\label{sec:router_analysis}
\begin{figure}[t]
    \centering
    \includegraphics[width=\columnwidth]
    {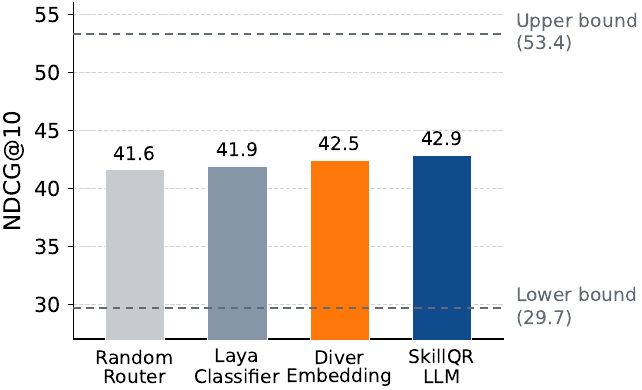}
    \caption{Router comparison on skill-sensitive BRIGHT queries. SkillQR’s LLM router achieves a modest numerical lead using only the rewriting LLM, without fine-tuning.}
    \label{fig:router_rank}
\end{figure}

We compare SkillQR's LLM router with Diver-based embedding routing, the non-autoregressive Laya classifier~\citep{convai2026laya}, and uniform random selection on BRIGHT, using Qwen2.5-7B-Instruct and Diver-Retriever with the full inference pipeline in Section~\ref{sec:skillqr_inference}. All routers choose from the same subject-specific skill set. For each query, we generate and evaluate a rewrite with every candidate skill. We retain 722 skill-sensitive queries, where skill choices yield different NDCG@10 scores, and evaluate each router by the retrieval performance of its selected skill's rewrite. Figure~\ref{fig:router_rank} shows that LLM routing achieves the highest observed score of 42.9 NDCG@10, closely matching Diver routing at 42.5 and scoring above Laya and random routing, without task-specific relevance labels or router fine-tuning. The retrospective upper bound selects the best skill for each query and reaches 53.4 NDCG@10. This gap motivates future work on lightweight router fine-tuning with synthetic query-skill pairs or in-context learning via examples pairing queries with their best-performing skills. Appendix~\ref{app:router_quality} details the routers in Figure~\ref{fig:router_rank} and explores online router improvement through an offline calibration study.
\subsection{Qualitative Analysis}
\label{sec:case study}
Figure~\ref{fig:skillqr_case} presents a case where SkillQR captures the relevant reasoning gap, while a generic rewrite instruction fails. The gold document is the definition of simple harmonic motion. Although the rewrite generated by the generic skill correctly identifies that answering the query requires knowledge of sampling theorems, it emphasizes solution articulation. In contrast, SkillQR understands that the corpus's concepts are most helpful to the query, and explicitly guides the model toward concept unfolding, leading the rewritten query to list relevant concepts such as the sampling theorem and Nyquist frequency. As a result, SkillQR achieves a substantially higher NDCG@10 of 43.1.
\section{Conclusion}
\label{Conclusion}
\vspace{-0.3 em}
We propose SkillQR, a training-free query rewriting framework that selects rewriting strategies from a skill set to align queries with corpus characteristics, achieving performance comparable to strong learned baselines while maintaining a favorable efficiency–effectiveness trade-off. SkillQR highlights the retrieval corpus as an overlooked source of rewriting signal, beyond stronger models or larger training sets. Our findings motivate corpus-aware strategy design for reason-intensive retrieval.

\section*{Limitations}
While SkillQR achieves competitive performance with strong efficiency, there are limitations. SkillQR relies on LLM-based proposer, judge, and rewriter agents, whose reasoning capabilities may be affected by pretraining biases, limitations in domain knowledge, and hallucinations due to long context. Our skill router analysis also shows a gap between the LLM router and the theoretical upper bound, suggesting that better query-to-skill matching could yield further gains in reason-intensive retrieval. Future versions of SkillQR may incorporate richer routing strategies and a dynamic skill refresh design for an evolving corpus. These improvements will contribute to a stronger system for reason-intensive retrieval.
\nocite{*}
\bibliography{custom}

\appendix

\section{Experiment Details}
\label{appendix:main_experiment}
\subsection{Models}
\label{appendix:model_details}
We list the detailed model backbones for all query rewriting methods used in our experiments in Table~\ref{tab:method_appendix}. We use a generation temperature of 0 for 7B-scale methods. For 14B-scale methods, we follow ThinkQE’s generation setup, using temperature 0.7 and concatenating two sampled responses with the original query for retrieval.
\begin{table}[ht]
\centering
\small
\setlength{\tabcolsep}{3pt}
\renewcommand{\arraystretch}{1.08}
\begin{tabularx}{\columnwidth}{lXl}
\toprule
\textbf{Method} & \textbf{Model Backbone} & \textbf{Source} \\
\midrule
ThinkQE &
DeepSeek-R1-Distill-Qwen-14B (QWEN-R1-Distill-14B) &
\cite{lei-etal-2025-thinkqe} \\
TongSearch &
TongSearch-QR-7B &
\cite{tongsearch} \\
Reasoner-Rewriter &
Reasoner-Rewriter-Qwen2.5-7B &
\cite{chen2025reasonembed} \\
SkillQR-7B &
Qwen2.5-7B-Instruct &
\cite{Yang2024Qwen25TR} \\
SkillQR-14B &
DeepSeek-R1-Distill-Qwen-14B &
\cite{deepseekai2025deepseekr1incentivizingreasoningcapability} \\
Diver-Retriever &
DIVER-Retriever-4B-1020 &
\cite{long2025diver} \\
Reason-Embed &
reason-embed-qwen3-4b-0928 &
\cite{chen2025reasonembed} \\
ReasonIR &
ReasonIR-8B &
\cite{shao2025reasonir} \\
\bottomrule
\end{tabularx}
\caption{Methods, backbone models, and implementation sources.}
\label{tab:method_appendix}
\end{table}

\subsection{Baseline Query Rewrite Prompts}
\label{appendix:baseline_prompts}
We discussed various query rewriting methods in this work. Here, we list the detailed prompt instructions used in ThinkQE, TongSearch, and Reasoner-Rewriter in Figure~\ref{fig:rewrite_prompt}.

\begin{figure*}[ht]
\centering
\begin{large}
\begin{footnotesize}
\begin{minipage}{\textwidth}
\begin{promptbox}{\textbf{TongSearch-QR}}
Instructions: \\
1. Identify the essential problem. \\
2. Think step by step to reason and describe what information could be relevant and helpful to address the questions in detail. \\
3. Draft an answer with as many thoughts as you have. \\
Query: \{original query\}
\end{promptbox}
\end{minipage}
\begin{minipage}{\textwidth}
\begin{promptbox}{\textbf{Reasoner-Rewriter (RRewriter)}}
Given a task and an input, first analyze the task and the input within the $<$think$>$ and $<$/think$>$ tags. In your analysis: \\
- Break down the requirements of the task \\
- Identify key components from the input \\
- Think step by step to reason about what should be included in the output \\
Then, within the $<$response$>$ and $<$/response$>$ tags, present the complete long output\\
\#\# Task\\
\{subject-specific task prompt\} \\
\#\# Input\\
\{original query\} 
\end{promptbox}
\end{minipage}
\begin{minipage}{\textwidth}
\begin{promptbox}{\textbf{ThinkQE}}
\begin{center}
Round 1 Instruction
\end{center}
Given a query and the provided passages (most of which may be incorrect or irrelevant), identify helpful information from the passages and use it to write a correct answering passage. Use your own knowledge, not just the example passages!\\
Query:
\{query\}\\
Possible helpful passages:
\{top\_passages\_str\}
\medskip
\hrule
\begin{center}
Round 2 Instruction
\end{center}
Given a query, the provided passages (most of which may be incorrect or irrelevant), and the previous round's answer, identify helpful information from the passages and refine the prior answer. \\
Ensure the output directly addresses the original query. Use your own knowledge, not just the example passages! \\
Query: 
\{query\}

Possible helpful passages: 
\{retrieved passage\}

Prior generated answer/revised query: 
\{round 1 generation\}
\end{promptbox}
\end{minipage}
\end{footnotesize}
\caption{Query Rewrite Instructions Used by Existing Query Rewriting Methods.}
\label{fig:rewrite_prompt}
\end{large}
\end{figure*}

\subsection{Human Skills for BRIGHT and R2MED}
\label{appendix:human}
We present the human skills for all subjects in Figure~\ref{fig:subject_prompt}. These skills are used by Oracle in Table~\ref{tab:oracle}, RRewriter in Tables ~\ref{tab:ndcg_bright} and ~\ref{tab:ndcg_r2med}, and Human Skills in Table~\ref{tab:skill_ablation}.
\begin{figure*}[t]
\centering
\begin{large}
\begin{footnotesize}
\begin{minipage}{\textwidth}
\begin{promptbox}{\textbf{Human Defined Subject Skills (Reasoner-Rewriter)}}
\textbf{StackExchange Subjects} \\
Generate information that is relevant and helpful to address the question in detail.

\vspace{0.3em}
\textbf{LeetCode} \\
Identify the required data structures and algorithms (e.g., DFS, DP, sorting, traversal), and provide a detailed explanation of how they are used to solve the problem.

\vspace{0.3em}
\textbf{Pony} \\
Generate the relevant Pony syntax and language features (including core components), and provide a detailed explanation of their usage.

\vspace{0.3em}
\textbf{AoPS} \\
Identify the mathematical functions and problem-solving techniques involved, and provide a detailed explanation of each.

\vspace{0.3em}
\textbf{TheoremQA\_theorems} \\
Generate the key theorem required to solve the problem, and provide a detailed explanation of the theorem.

\vspace{0.3em}
\textbf{TheoremQA\_questions} \\
Identify the mathematical concepts, formulas, and theorems required, and provide detailed explanations for each.

\vspace{0.3em}
\textbf{All R2MED Subjects} \\
Generate information that is relevant and helpful to address the question in detail.
\end{promptbox}
\end{minipage}
\end{footnotesize}
\caption{Human-Defined Rewrite Strategies for Subjects in BRIGHT and R2MED.}
\label{fig:subject_prompt}
\end{large}
\end{figure*}
\subsection{SkillQR Prompts}
\label{appendix: skillqr_prompts}
\begin{figure*}[t]
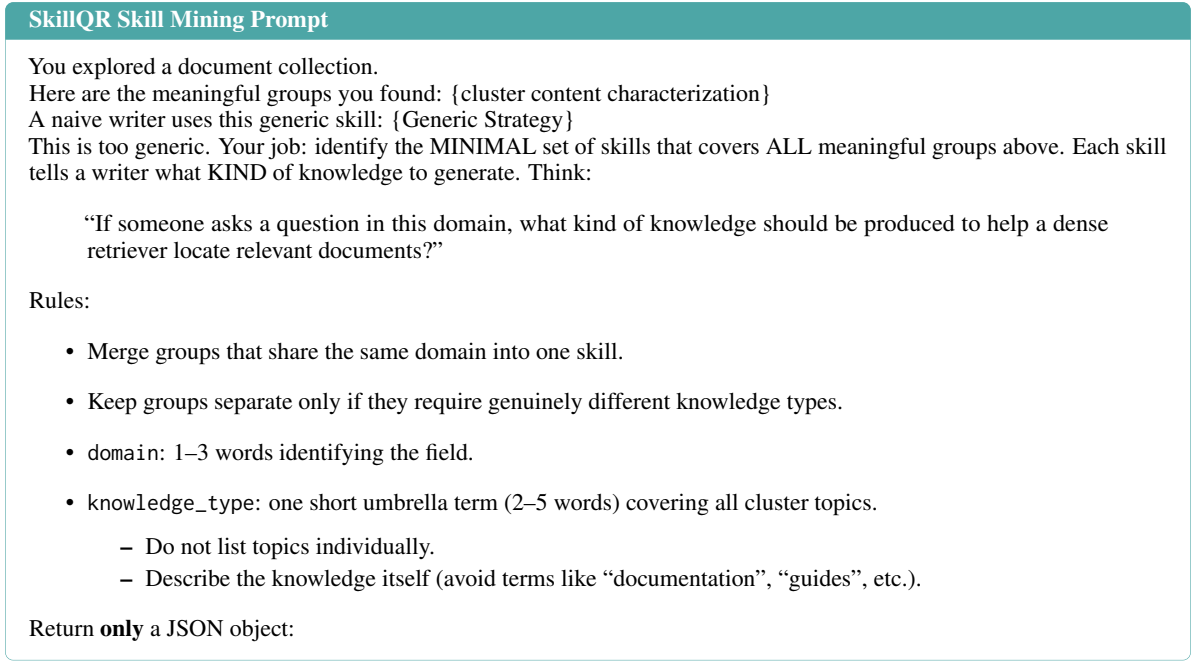

\centering
\begin{minipage}{0.98\textwidth}
\footnotesize
\begin{promptbox}{\textbf{SkillQR Skill Mining Prompt}}
You explored a document collection.

Here are the meaningful groups you found:
\{cluster content characterization\}

A naive writer uses this generic skill: \{Generic Strategy\}

This is too generic. Your job: identify the MINIMAL set of skills that covers ALL meaningful groups above.
Each skill tells a writer what KIND of knowledge to generate. Think:
\begin{quote}
“If someone asks a question in this domain, what kind of knowledge should be produced
to help a dense retriever locate relevant documents?”
\end{quote}
Rules:
\begin{itemize}
    \item Merge groups that share the same domain into one skill.
    \item Keep groups separate only if they require genuinely different knowledge types.
    \item \texttt{domain}: 1--3 words identifying the field.
    \item \texttt{knowledge\_type}: one short umbrella term (2--5 words) covering all cluster topics.
    \begin{itemize}
        \item Do not list topics individually.
        \item Describe the knowledge itself (avoid terms like ``documentation'', ``guides'', etc.).
    \end{itemize}
\end{itemize}

Return \textbf{only} a JSON object:
\end{promptbox}
\end{minipage}
\caption{Skill mining prompt used in SkillQR to extract a minimal set of domain-specific knowledge types from clustered documents.}
\label{fig:skill_mining_prompt}
\end{figure*}

\begin{figure*}[t]
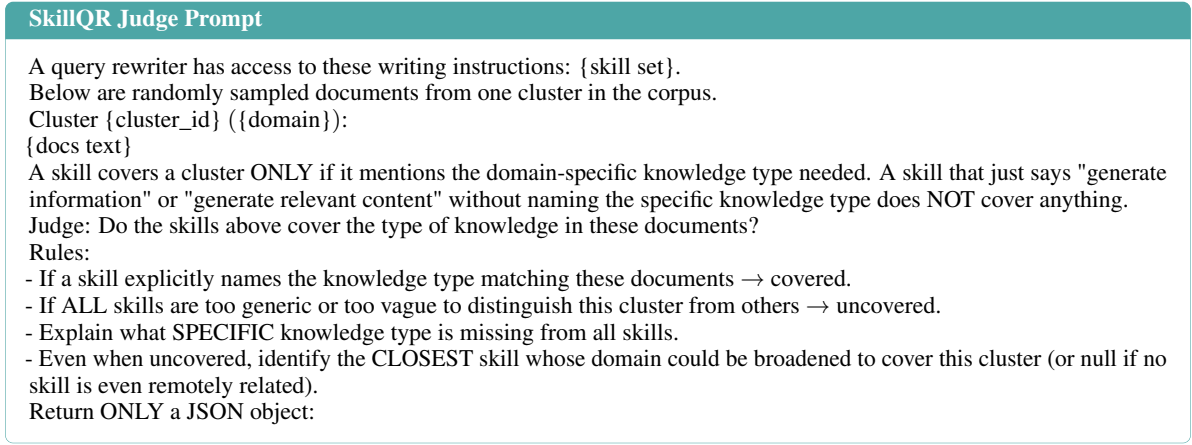

\centering
\begin{minipage}{0.98\textwidth}
\footnotesize
\begin{promptbox}{\textbf{SkillQR Judge Prompt}}
A query rewriter has access to these writing instructions: \{skill set\}.\\
Below are randomly sampled documents from one cluster in the corpus.\\
Cluster \{cluster\_id\} $($\{domain\}$)$: \\
\{docs text\}\\
A skill covers a cluster ONLY if it mentions the domain-specific knowledge type
needed. A skill that just says "generate information" or "generate relevant content" without
naming the specific knowledge type does NOT cover anything.

Judge: Do the skills above cover the type of knowledge in these documents?\\
Rules:\\
- If a skill explicitly names the knowledge type matching these documents → covered.\\
- If ALL skills are too generic or too vague to distinguish this cluster from others → uncovered.\\
- Explain what SPECIFIC knowledge type is missing from all skills.\\
- Even when uncovered, identify the CLOSEST skill whose domain could be broadened to cover this cluster (or null if no skill is even remotely related).\\
Return ONLY a JSON object:
\end{promptbox}
\end{minipage}
\caption{SkillQR Judge prompt to evaluate the quality of the skillsets.}
\label{fig:skill_judge_prompt}
\end{figure*}

Figure~\ref{fig:skill_mining_prompt} presents the proposer agent’s full prompt for skill mining, while Figure~\ref{fig:skill_judge_prompt} shows the judge agent’s prompts for skill set evaluation. Figure~\ref{fig:skill_refine_prompt} presents the proposer's prompt for refining skill sets. The rewriter agent’s full routing and generation prompt is provided in Figure~\ref{fig:skillqr_rewrite_prompt}.
\begin{figure*}[t]
\centering
\footnotesize
\begin{minipage}{0.97\textwidth}
\begin{promptbox}{SkillQR Rewrite Generation Prompt}

\textbf{Routing Prompt}
Given a query and a numbered list of writing strategies, choose the single most fitting strategy. Do not answer the query.\\
Query: \{original query\} \\
Available writing strategies: \{discovered skills\} \\
Think step by step:
\begin{itemize}
    \item What is the genre of this query?
    \item What knowledge is required to solve this query?
    \item Which strategy description matches the genre and knowledge?
    \item If a domain-specific strategy fits, prefer it over a general-purpose one.
\end{itemize}
Pick a strategy for the query.\\
Output format: \\
Reasoning: [reasoning] \\
Select: Strategy N
\medskip
\hrule
\medskip
\textbf{Rewrite}
Given a task and an input, first analyze them within the \texttt{<think>} and \texttt{</think>} tags. In your analysis:
\begin{itemize}
    \item Break down the requirements of the task;
    \item Identify key components from the input;
    \item Reason step by step about what should be included in the output.
\end{itemize}
Then, within the \texttt{<response>} and \texttt{</response>} tags, present the complete output. \\
\noindent\textbf{Task}\\
\{selected skill\} \\
\noindent\textbf{Input} \\
\{original query\}\\
\noindent\textbf{Possible helpful passages} \\
\{top-$k$ documents retrieved by the dense retriever using the original query\}
\end{promptbox}
\end{minipage}
\caption{Routing and rewriting prompt in SkillQR.}
\label{fig:skillqr_rewrite_prompt}
\end{figure*}
\begin{figure*}[t]
\centering
\begin{minipage}{0.98\textwidth}
\footnotesize
\begin{promptbox}{\textbf{SkillQR Refine Prompt}}
The judge found these clusters are close to skill \{skill id\} domain but NOT covered: \{uncovered clusters\}. Refine the instruction to also cover these clusters.\\
Keep the original domain focus $-$ just broaden the \{content-type\} slightly. \\
Answer: \\
1. domain: The domain of this skill (1-3 words, keep the same or broaden slightly) \\
2. knowledge\_type: Find ONE short umbrella term that covers both the original scope AND the uncovered clusters (2-5 words, no topic lists)\\
Return ONLY a JSON object:
\medskip
\hrule
\medskip
The following clusters are NOT covered by any existing skill: \{uncovered content-type\}.\\
Your job: create a NEW skill instruction that covers the uncovered clusters above.
The new skill must NOT duplicate what existing skills already cover.
\end{promptbox}
\end{minipage}
\caption{SkillQR Proposer refine prompt to update skillsets. The top section is for updating an existing skill. The bottom section is to add a new skill.}
\label{fig:skill_refine_prompt}
\end{figure*}
\section{Time Analysis}
\label{app:time_analysis}
Table~\ref{tab:offline_time} reports offline skill-mining times and corpus sizes for BRIGHT subjects. Timings exclude embedding generation, clustering, and model initialization. Most subjects complete skill mining within 10 minutes.
\begin{table}[h]
\centering
\setlength{\tabcolsep}{6pt}
\footnotesize
\begin{tabular}{lrr}
\toprule
\textbf{Subject} & \textbf{Size} & \textbf{Time (min)} \\
\midrule
Bio.    & 57,359  & 7.2 \\
Earth.  & 121,249 & 9.5 \\
Econ.   & 50,220  & 7.3 \\
Psy.    & 52,835  & 7.4 \\
Rob.    & 61,961  & 6.6 \\
Stack.  & 107,081 & 12.7 \\
Sus.    & 60,792  & 8.3 \\
Leet.   & 413,932 & 21.0 \\
Pony    & 7,894   & 4.4 \\
Aops    & 188,002 & 16.2 \\
TheoQ.  & 188,002 & 16.5 \\
TheoT.  & 23,839  & 8.4 \\
\bottomrule
\end{tabular}
\caption{Offline skill mining time on BRIGHT subjects using Diver-Retriever embeddings.}
\label{tab:offline_time}
\end{table}
We also provide a subject breakdown of inference time for all methods in Table~\ref{tab:cost_comparison}.
\begin{table*}[t]
\centering
\setlength{\tabcolsep}{1.9pt}
\renewcommand{\arraystretch}{1.05}
\footnotesize
\begin{tabular}{lccccccccccccc}
\toprule
& &
\multicolumn{7}{c}{\textbf{StackExchange}} &
\multicolumn{2}{c}{\textbf{Coding}} &
\multicolumn{3}{c}{\textbf{Theorem-based}} \\
\cmidrule(r{6pt}){3-9} \cmidrule(r{6pt}){10-11} \cmidrule(r{6pt}l{6pt}){12-14}
\textbf{Method} & \textbf{Avg} &
\textbf{Bio.} & \textbf{Earth.} & \textbf{Econ.} & \textbf{Psy.} &
\textbf{Rob.} & \textbf{Stack.} & \textbf{Sus.} &
\textbf{Leet.} & \textbf{Pony} & \textbf{Aops} &
\textbf{TheoQ.} & \textbf{TheoT.} \\
\midrule
ThinkQE-14B & 85.3 & 33.7 & 34.3 & 43.6 & 31.1 & 50.6 & 65.9 & 40.6 & 234.3 & 109.8 & 234.5 & 61.3 & 83.4 \\
SkillQR-14B & 38.1 & 20.5 & 24.2 & 26.1 & 23.8 & 26.1 & 35.4 & 23.9 & 69.9 & 51.7 & 70.8 & 44.8 & 40.0 \\
TongSearch-7B & 17.7 & 18.5 & 14.2 & 17.5 & 22.5 & 22.9 & 13.3 & 14.4 & 24.4 & 25.2 & 15.6 & 9.5 & 14.6 \\
RRewriter-7B & 21.1 & 14.7 & 15.9 & 20.1 & 16.2 & 21.3 & 27.0 & 18.4 & 19.2 & 15.8 & 34.4 & 24.7 & 24.9 \\
SkillQR-7B & 13.9 & 12.2 & 14.2 & 13.2 & 12.6 & 14.3 & 15.5 & 14.0 & 11.9 & 14.3 & 15.8 & 14.5 & 14.1 \\
\bottomrule
\end{tabular}
\caption{Per-query inference time (seconds) on BRIGHT across methods. Averages are computed over 240 queries (20 per subject).}
\label{tab:cost_comparison}
\end{table*}
\begin{table}[t]
\centering
\small
\setlength{\tabcolsep}{8pt}
\renewcommand{\arraystretch}{1.1}
\begin{tabular}{crrr}
\toprule
$k$ & LLM & Calibrated & Gain \\
\midrule
10 & 42.90 & 42.52 & $-0.38$ \\
20 & 42.85 & 42.94 & $+0.09$ \\
30 & 42.80 & 43.27 & $+0.47$ \\
\bottomrule
\end{tabular}
\caption{Held-out subject-averaged NDCG@10 after calibration
with $k$ labeled queries per subject.
Results average over 1,000 random splits.}
\label{tab:router_calibration}
\end{table}
\section{Additional Design Choice Discussion}
\subsection{Skill Discovery Design and Sensitivity}
\label{app:cluster_ablation}
SkillQR adopts MiniBatchKMeans for its speed and stability on large corpora. We initially tested Birch clustering, but found it produced a single dominant cluster alongside many small fragments on retriever embeddings, leaving the proposer agent without a useful corpus overview. Re-clustering the dominant cluster with MiniBatchKMeans was required as a follow-up step after Birch, so we adopt MiniBatchKMeans directly.

We evaluate skill discovery sensitivity on Biology, Pony, and TheoremQA\_theorems, covering the StackExchange, coding, and theorem-based task groups in BRIGHT. Using Diver-Retriever and Qwen2.5-7B-Instruct with five PRF passages, we independently vary the cluster count $K$ and skill cap $N_S$ while retaining the remaining hyperparameters.

As shown in Table~\ref{tab:discovery_sensitivity}, halving or doubling the cluster count reduces averaged NDCG@10 by only 0.84 and 1.30 points, respectively. Despite a fourfold difference between the smallest and largest cluster counts, retrieval performance remains relatively stable, suggesting robustness to the initial clustering granularity within the evaluated range. We believe that iterative proposer-judge refinement contributes to this stability: by identifying uncovered knowledge types and broadening skill coverage, the refinement process may reduce dependence on the initial partition. Increasing the skill cap to ten reduces the mean by 2.62 points, although Biology improves, indicating that a larger skill set does not consistently improve retrieval. 
\begin{table*}[t]
\centering
\small
\begin{tabular}{lccrrrr}
\toprule
Configuration & Cluster count $K$ & Skill cap $N_S$
& Biology & Pony & TheoT & Avg. \\
\midrule
Default
& $\sqrt{D}$ & 5
& 51.90 & \textbf{22.75} & \textbf{46.92} & \textbf{40.52} \\
Coarser clustering
& $0.5\sqrt{D}$ & 5
& 52.31 & 21.61 & 45.11 & 39.68 \\
Finer clustering
& $2\sqrt{D}$ & 5
& \textbf{53.12} & 20.68 & 43.87 & 39.22 \\
Larger skill cap
& $\sqrt{D}$ & 10
& 52.96 & 16.96 & 43.79 & 37.90 \\
\bottomrule
\end{tabular}
\caption{
NDCG@10 sensitivity to cluster count and discovered-skill cap using
Diver-Retriever and Qwen2.5-7B-Instruct with five PRF passages.
Default results are reused from the existing evaluation.
Best results per column are bolded. $D$ represents the size of the retrieval corpus. 
}
\label{tab:discovery_sensitivity}
\end{table*}
\subsection{Router Analysis}
\label{app:router_quality}
\subsubsection{Implementation Details}
In Section~\ref{sec:router_analysis}, the router comparison uses the same saved Qwen2.5-7B-Instruct rewrites and Diver-retrieved PRF documents. We retain the 722 queries for which candidate skills yield different NDCG@10 scores, averaging NDCG@10 within each subject and then equally across subjects. All routers select from the generic instruction and the discovered skill set. Random routing reports the exact expectation under uniform selection over distinct skill instructions. The embedding router selects the skill whose Diver embedding has the highest cosine similarity to the original query. We use Laya~\citep{convai2026laya} in a zero-shot way by providing the original query, subject, and candidate skill instructions as a multiple-choice selection task and choosing the highest-probability option. No baseline receives document-relevance judgments or candidate retrieval scores as input.
\subsubsection{Potential for Router Improvement}
\begin{figure*}[t]
\centering
\includegraphics[width=\textwidth]{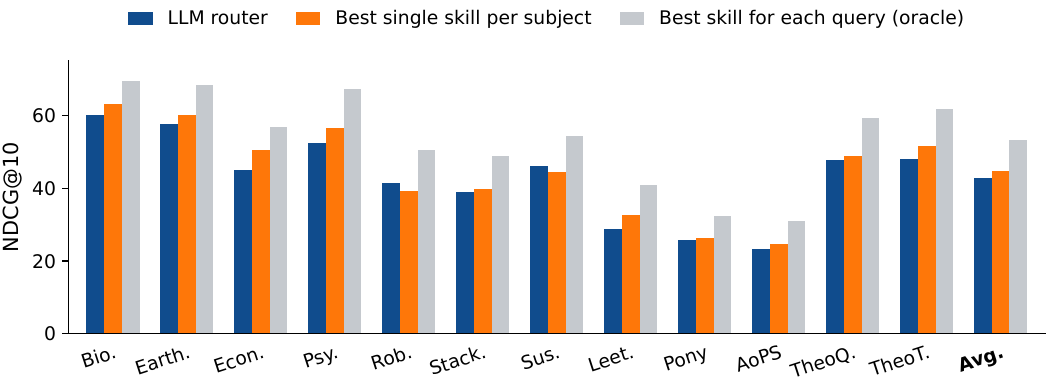}
\caption{Skill selection NDCG@10 on skill-sensitive BRIGHT
queries with Diver-Retriever and Qwen2.5-7B-Instruct.
Retrospective best-skill selection highlights further
improvement potential for routing.}
\label{fig:router_quality_appendix}
\end{figure*}

We examine further potential improvement in skill routing. Figure~\ref{fig:router_quality_appendix} compares LLM routing with two retrospective reference policies: selecting the best single skill per subject and selecting the best skill for each query (oracle). Both use ground-truth retrieval scores to select skills. On the same 722 skill-sensitive queries, selecting the \textbf{best single skill per subject} achieves 44.79 NDCG@10, compared with 42.90 for LLM routing. This gap motivates feedback-based adaptation during deployment, where accumulated retrieval feedback could identify high-performing skills and guide subsequent skill selection.

To explore the potential for online router adaptation, we conduct an offline calibration replay using gold relevance judgments to score all candidate skills on a subset of queries. For each subject, we select the skill with the highest mean calibration NDCG@10 and apply it to disjoint held-out queries, without updating model parameters. Table~\ref{tab:router_calibration} reports subject-averaged results over 1,000 random splits. With 30 calibration queries per subject, calibrated selection achieves 43.27 NDCG@10 versus 42.80 for LLM routing on the same held-out queries, a modest 0.47-point gain. These exploratory results suggest that accumulated feedback could improve skill selection during deployment. The replay assumes gold relevance judgments and complete feedback for all candidate skills, which means efficient feedback collection and continual online adaptation remain valuable future work. Such feedback-based extensions introduce additional supervision, whereas the current SkillQR pipeline requires no task-specific relevance labels or router fine-tuning.
\subsection{Skill Performance}
\label{app: skill_effectiveness}
Table~\ref{tab:skill_ablation_reason_embed} shows that SkillQR's discovered skills also work with Reason-Embed, matching the skill quality trend in Section~\ref{sec:skill_effectiveness_ablation}. We also evaluate an alternative skill generation strategy, SkillQR-Syn, which skips the proposer-judge aggregation. SkillQR-Syn mines skills from each cluster independently to form a skill pool. At inference time, the retriever selects the top-5 skills most semantically similar to the query, and the rewriter synthesizes a query-specific skill from these candidates. This produces a finer-grained, query-tailored alternative to SkillQR’s compact, corpus-level skill set. Table~\ref{tab:skill_ablation_reason_embed} shows that SkillQR-Syn underperforms SkillQR, suggesting that online synthesis from local skill fragments lacks the holistic corpus view produced by the dual-agent refinement loop.
\begin{table*}[t]
\centering
\label{tab:sheet_style}
\setlength{\tabcolsep}{4.0 pt}
\renewcommand{\arraystretch}{1.05}
\footnotesize
\begin{tabular}{lccccccccccccc}
\toprule
& &
\multicolumn{7}{c}{\textbf{StackExchange}} &
\multicolumn{2}{c}{\textbf{Coding}} &
\multicolumn{3}{c}{\textbf{Theorem-based}} \\
\cmidrule(r{6pt}){3-9} \cmidrule(r{6pt}){10-11} \cmidrule(r{6pt}l{6pt}){12-14}
\textbf{Method} & \textbf{Avg} &
\textbf{Bio.} & \textbf{Earth.} & \textbf{Econ.} & \textbf{Psy.} &
\textbf{Rob.} & \textbf{Stack.} & \textbf{Sus.} &
\textbf{Leet.} & \textbf{Pony} & \textbf{Aops} &
\textbf{TheoQ.} & \textbf{TheoT.} \\
\midrule
Original &36.2 & 51.8 & 52.5 & 33.5 & 44.8 & 34.3 & 34.6 & 35.7 & 38.2 & 9.3 & 12.8 & 40.8 & 46.3\\
Generic & 36.8 & 51.9 & 53.8 & 30.4 & 47.1 & 31.4 & 34.9 & 37.2 & 34.4 & 24.7 & 13.6 & 41.8 & 40.5\\
Human & 38.1 & 51.8 & 54.0 & 30.4 & 47.1 & 31.3 & 34.9 & 37.0 & 33.1 & 31.7 & 14.5 & 43.3 & 48.0 \\
SkillQR & 37.5 & 52.2 & 57.2 & 31.4 & 45.1 & 31.8 & 36.1 & 33.7 & 34.0 & 25.8 & 14.0 & 42.2 & 46.8 \\
SkillQR-Syn & 36.3 & 52.7 & 51.9 & 29.8 & 43.2 & 31.4 & 36.8 & 30.9 & 34.6 & 24.9 & 13.9 & 41.0 & 44.7\\
\bottomrule
\end{tabular}
\caption{Skill quality NDCG@10 comparison on BRIGHT with Reason-Embed and Qwen2.5-7B-Instruct rewriter. SkillQR methods in this table do not inherently include PRF.}
\label{tab:skill_ablation_reason_embed}
\end{table*}
\section{Additional Qualitative Analysis}
\label{sec:additional_qualitative}
\begin{figure*}[t]
    \includegraphics[width=1.0\textwidth]{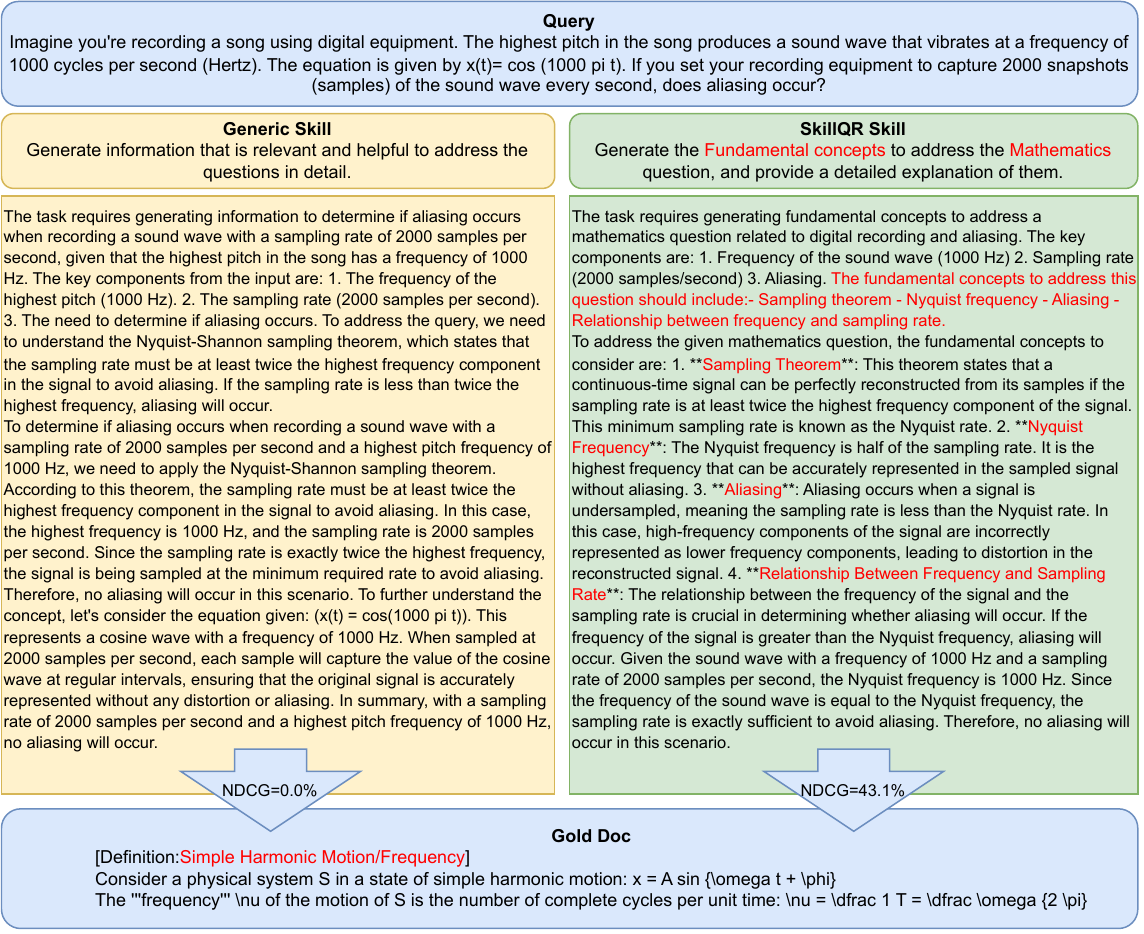}
    \vspace{-1.0em}
    \caption{SkillQR directs the focus of the rewrite from problem-solving into theorem-explaining, improving the NDCG@10 score from 0.0 to 43.1.}
    \label{fig:skillqr_case_full}
\end{figure*}

We include the complete rewrite case study in Section~\ref{sec:case study} in Figure~\ref{fig:skillqr_case_full}. We can see that the generic skill focuses on generating a solution to the query at hand via listing numbers and formulas, while SkillQR realizes that the core reasoning need is concept elicitation.
\section{AI Use Statement}
Closed-source LLMs were used only to improve the clarity and presentation of the manuscript. They did not contribute to the research design, experiments, analysis, or conclusions, for which the authors take full responsibility.
\end{document}